\documentclass[aps,prl,reprint,showpacs,superscriptaddress]{revtex4-2}

\usepackage{amsmath}
\usepackage{graphicx}
\usepackage{lmodern}
\usepackage{amsmath}
\usepackage{color}
\usepackage{amssymb}
\usepackage{bm}
\usepackage{dsfont}
\usepackage{braket}
\usepackage{mathtools}
\usepackage{afterpage}

\newcommand{\br}{\bm{r}}
 
\newcommand{\bR}{\bm{R}}
\newcommand{\bu}{\bm{u}}
\newcommand{\bx}{\bm{x}}

\newcommand{\bk}{\bm{k}}

\newcommand{\bK}{\bm{K}}

\newcommand{\bp}{\bm{p}}
\newcommand{\bq}{\bm{q}}

\newcommand{\bQ}{\bm{Q}}

\newcommand{\be}{\begin{equation}}
\newcommand{\ee}{\end{equation}}

\usepackage[dvipsnames]{xcolor}
\usepackage[colorlinks=true]{hyperref}
\hypersetup{
    colorlinks=true,
    linkcolor=blue,
    citecolor=blue,
    urlcolor=blue
} 

\makeatletter
\def\maketitle{
\@author@finish
\title@column\titleblock@produce
\suppressfloats[t]}
\makeatother

\begin{document}
\title{Enhanced Diffusion from Twist-Angle Disorder}

\author{Nicole S. Ticea}
\email{nticea@stanford.edu}
\affiliation{Department of Applied Physics, Stanford University, Stanford, CA 94305, USA}
\affiliation{Google Quantum AI, Goleta, CA, 93111, USA}

\author{Yi-Ming Wu}
\email{yimwu@zju.edu.cn}
\affiliation{Institute for Advanced Study in Physics, Zhejiang University, Hangzhou 310027, China}

\date{\today}

\begin{abstract}
  We consider the effect of twist-angle disorder on the single-particle properties of twisted bilayer systems. These materials are a natural playground for studying how tiny, uncontrolled spatial variations may become unusually consequential; a local change in the twist geometry can give rise to spatially-correlated alterations of the parent Hamiltonian, which in turn affect the coarse-grained properties of the substrate such as transport and ordering tendencies. In analogy to electrodynamics, where the magnetic field is given by the curl of the vector potential, we model the twist angle as the curl of the relative displacement between the two bilayers. We find that geometric scattering constraints imposed here by the twist structure substantially suppress momentum relaxation relative to mean-rate-matched white noise. We quantify these effects and show that the problem can be mapped onto a non-linear sigma model (NLSM). 
  
\end{abstract}
\maketitle

\paragraph{Introduction.} Disorder strength alone does not determine localisation and transport. Consider, for example, the case of disordered fermions with spin orbit coupling, which display a genuine metallic phase and Anderson transition \cite{10.1143/PTP.63.707,PhysRevB.40.5325,RevModPhys.80.1355,PhysRevLett.89.256601}. This example underscores how coupling the disorder to an operator other than the density can substantially alter the fate of single-particle localisation in two dimensions. We consider a similar situation here, modelling twist-angle disorder as deviations in the relative displacement between two coupled bilayers. The intrinsic relation between twist angle and displacement mirrors the relationship between magnetic field and vector potential. This sets up a natural parallel with the problem of a charged particle in a static random magnetic field \cite{PhysRevB.49.16609}, where it was found that---while the problem can be mapped onto a NLSM in the unitary class, and the particles therefore eventually localise in $2d$---the intrinsic symmetries can result in an enormous increase in the localisation length. A similar enhancement of the diffusivity is present in the twist-angle disorder case. We believe this to be, in part, attributable to the intricate structure of the scattering kernel---inherited from the fact that we are considering deviations from a rigid twist, which can dispatch correlations across large scales. 

We are particularly motivated to study the effect of twist-angle disorder on the phase diagrams of popular bilayer systems, such as in twisted bilayer graphene (TBLG) and transition metal dichalcogenides (TMDs) \cite{zeldov,PhysRevResearch.3.013153,Kazmierczak2021,SimonTurkel,deJong2022,bathen2025precisetwistangledetermination}. For example, the pairing symmetry of superconductivity in TBLG and twisted bilayer WSe$_2$ and MoSe$_2$ is unknown \cite{Oh2021,Isobe2018,Wu2018,XuBalents2018,LiuPairing2018,Kennes2018,Gonzalez2019,Samajdar2020,Chou2021,Wu2023,SchradeFu2024,Chen2026,Xia2025,Guo2025,zeldov}. If disorder in these systems has a strong localising effect, then conventional $s$-wave pairing symmetry would be less affected by nonmagnetic disorder \cite{Anderson1959}. Fractional quantum Hall states, which are sensitive to inhomogeneities \cite{MacDonald1986,Sheng2003,Deng2014,WangDisorder2018,Zhu2019}, could likewise be strongly affected by twist-angle disorder. This has an important bearing on recent observations of fractional Chern insulators in twisted bilayer MoTe$_2$ \cite{Cai2023,Zeng2023,Park2023,Xu2023,Wang2024,Reddy2023,Yu2024,MoralesDuran2024,Jia2024,Redekop2024,PhysRevB.109.115111,PhysRevLett.133.186602,3crp-gb3d}. 

This work builds on prior efforts to characterise disorder in twisted compounds, such as attempts based on the Anderson model \cite{,PhysRevLett.134.126301,9lc1-9m8t}, spatially-random Fermi velocities \cite{PhysRevResearch.2.043416}, random gauge fields \cite{c6rt-6qg1}, and nonuniform lattice distortions \cite{PhysRevB.105.245408}. The effects of twist-angle disorder on electronic transport have also been studied with tight-binding simulations \cite{PhysRevResearch.2.023325,ciepielewski2024transporteffectstwistangledisorder}. Despite these advances, a simple model capturing the essential nature of twist-angle disorder is still lacking.

\begin{figure*}[t] 
    \centering
    \includegraphics[width=\textwidth]{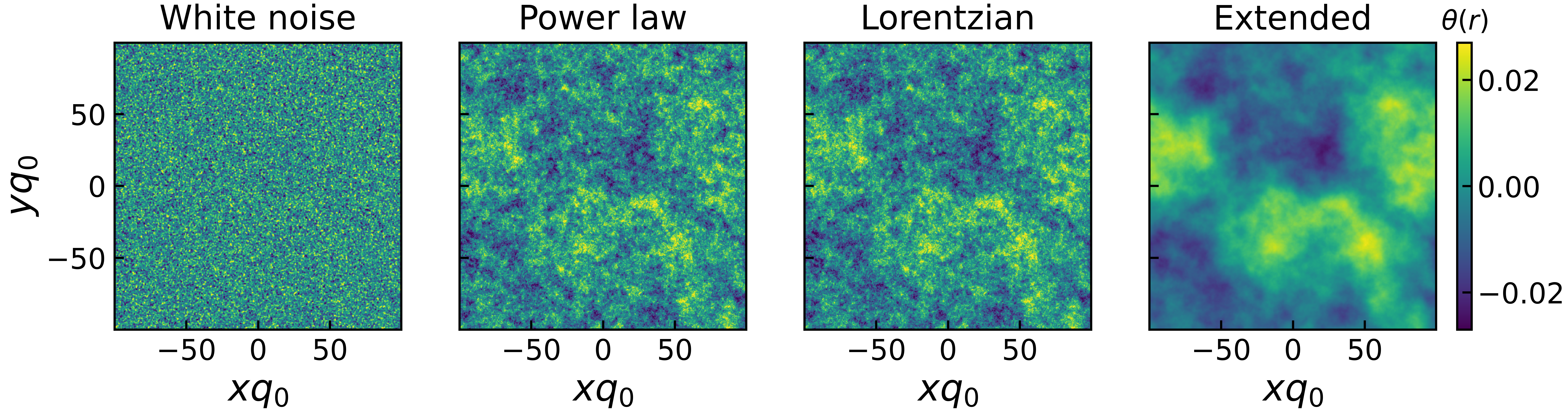} 
    \vspace{5mm}
    \includegraphics[width=\textwidth]{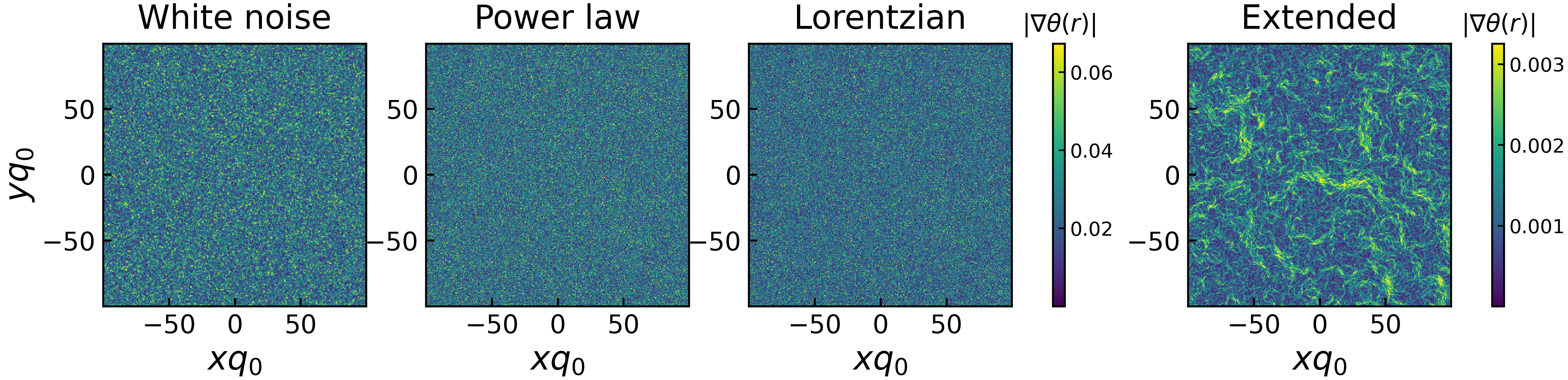} 
    \caption{Top row: sampling $\theta(\br)$ using four different covariance kernels: uncorrelated disorder, power-law-correlated disorder ($\alpha=2$), Lorentzian disorder ($\xi=20$), and disorder generated by a kernel with quartic momentum contributions ($\alpha=0$, $\xi=20$). Bottom row: $|\nabla\theta(\br)|$ for the same covariance kernels.}
    \label{fig:all_correlators}
\end{figure*}

\paragraph{Disorder model.} While twist-angle disorder is most commonly referenced in the context of moiré materials, the phenomenon exists independently of any particular system. For this reason, we will begin with a model-agnostic discussion of how to model a rigid twist and deviations thereof \cite{Balents_2019}. Let $\bR$ denote a point in the undeformed solid. Upon elastic deformation, $\bR \to \br$. We introduce the displacement field, $\bu(\br)$, relating the deformed and undeformed coordinates via $\br = \bR + \bu(\br)$. Note that we are working in `Eulerian' coordinates, where the displacement field is associated with an actual location in the \textit{deformed} solid $\br$, rather than a point in the original lattice $\bR$ \footnote{The latter prescription, aka using `Lagrangian' coordinates, can be problematic when the point $\br$ no longer corresponds to the location of a point in physical space.}. In the case of twisted bilayer compounds, the displacement field can be made to account for a rigid twist by angle $\theta$:
\begin{equation}
    \bu_\ell(\br) = (-1)^\ell\frac{\theta}{2}\hat{\bm{z}}\times\br \label{eq:u}
\end{equation}
Now let us suppose that the twist angle is subject to some disorder such that it is no longer uniform. Naively, one would think to promote $\theta \to \theta(\br) \equiv \theta_0 + \delta\theta(\br)$. But there is a problem with this prescription: disorder at points far away from the origin result in larger displacements. This is inconsistent with the observed phenomenology of twisted compounds, where mesoscopic patches of the material are at a consistent uniform twist angle \cite{zeldov}. Instead, we decompose the displacement into a rigid twist and deviations thereof:
\begin{equation}
    \bu_\ell(\br) = (-1)^\ell\frac{\theta}{2}\hat{z}\times\br + \tilde{\bu}_\ell(\br) \label{eq:u_disorder}
\end{equation}
The disorder piece satisfies a local rotational constraint,
\begin{equation}
    \frac{1}{2}\nabla\times \tilde{\bu}_\ell(\br) = \delta\theta_\ell(\br) \hat{z} \label{eq:curl_tilde_u},
\end{equation}
Since only the relative twist angle matters, we can introduce $\tilde{\bm{u}}=\tilde{\bm{u}}_1-\tilde{\bm{u}}_2$ (and ${\bm{u}}={\bm{u}}_1-{\bm{u}}_2$), which is similarly related to the relative twist angle $\delta\theta(\br)=\delta\theta_1(\br)-\delta\theta_2(\br)$. Taking the curl of both sides of Eq. \eqref{eq:curl_tilde_u}, we can reconstruct $\tilde{\bu}$:
\begin{equation}
    \tilde{\bu}(\br) = \nabla\phi(\br) + \frac{1}{\pi}\int d^2 \br' \delta\theta(\br')\frac{\hat{z}\times (\br-\br')}{|\br-\br'|^2}
\end{equation}
The first term $\nabla\phi$, does not contribute to the twist (it represents strain), and so we neglect it for now. We posit that the twist-angle disorder has zero mean and some correlation kernel $C_\theta(|\br-\br'|) \equiv \langle \delta\theta(\br)\delta\theta(\br')\rangle$. For example, consider white-noise ($C_\theta(|\br|)\sim \delta^2(|\br|)$), power-law ($C_\theta(|\br|)\sim \br^{-\zeta}$), and Lorentz-correlated disorder ($S_\theta(k)\sim 1/(1+\xi^2k^2)$), for some parameters $\zeta$ and $\xi$. Plotting $\theta(\br)$ and $|\nabla \theta(\br)|$ for these three disorder configurations in Fig. \ref{fig:all_correlators}, we see that domain walls begin to appear for the power-law and Lorentz-correlated disorder. However, the domain walls---a key experimental observation \cite{zeldov}---are poorly resolved. We can tweak the power spectrum of the kernel to sharpen the gradient. For example, consider the following `extended' power spectrum,
\begin{equation}
    S_\theta(\bk) = \frac{1}{(1+\xi^2 k^2)^\alpha},
\end{equation}
which is related to the correlation function in real space, $C_\theta$, by a Fourier transform. The twist angle in space, as well as its gradient, is plotted in the rightmost panel of Fig. \ref{fig:all_correlators} for $\alpha=2$. The spatial distribution of $\theta(\br)$ now matches that of experiment, see Ref. \cite{zeldov}. In what follows, we will consider an arbitrary kernel $S(\bk)$ and find the generic solution for the disorder-averaged case. We can then easily specialise to the various kernels used to produce the panels in Fig. \ref{fig:all_correlators}.

\paragraph{Moire model.} We will now describe how twist angles and disorder enter into the formulation of twisted TMDs. For convenience of discussion, we consider the moiré band structure from twisted heterobilayer transition metal dichalcogenide, e.g. WSe$_2$/MoSe$_2$ \cite{PhysRevLett.121.026402}. Since both layers have almost the same lattice constant, the moiré pattern resulting from lattice mismatch can be neglected. Moreover, these TMDs have a large, valley-contrasting spin-orbit coupling in the valence bands; at energies close to the valence band top, spin-valley locking effectively leads to a spinless fermion per valley for the moiré band structures. Lastly, the energy difference between bands deriving from the top and bottom layers makes it possible to tune the Fermi level such that it crosses the valence band of one layer but remains within the gap of the other. Therefore, the resulting low energy moiré bands for each valley in the `clean' limit (with no angle disorder) can be effectively described by the following one-band continuum limit Hamiltonian
\begin{equation}
     \begin{aligned}
         H&=\varepsilon(\bk)+\Delta(\br),\\
     \Delta(\br) &= \frac{V}{2}e^{i\varphi} \sum_\mu e^{-i\bQ_\mu\cdot [\bu_1(\br)-\bu_2(\br)]} + h.c.
     \end{aligned}\label{eq:H}
 \end{equation} 
where $\varepsilon(\bk)$ is the free-particle dispersion, $V>0$ is the potential strength, $\varphi$ is a phase factor, and $\{\bm{Q}_{\mu}\}_{\mu=1}^3$ are three atomic reciprocal lattice vectors related to each other by $\pm\frac{2\pi}{3}$ rotation, as described in Ref. \cite{Balents_2019}. This form is constrained by the fact that energy depends only on twist angle, and must be therefore invariant under a shift of Bravais lattice vector on one of the two layers. In the presence of angle disorder this translation invariance is broken, but for small twist-angle disorder we make the approximation that the form $\Delta(\br)$ remains valid; disorder effects enter only through $\bm{u}_\ell(\br)$. Substituting Eq. \ref{eq:u_disorder} into the exponential for $\Delta(\br)$, 
for the exponential we obtain $\bQ_\mu\cdot\bu(\br) = \theta\bQ_\mu\cdot\hat{z}\times\br + \bQ_\mu\cdot\tilde{\bu}(\br)$. Defining $\theta\bQ_\mu\cdot\hat{z}\times\br = (-\theta\hat{z}\times\bQ_\mu)\cdot\br \equiv \bq_\mu \cdot\br$, it then follows that 
\begin{equation}
    \begin{aligned}
        &\bQ_\mu\cdot\bu(\br) = \bq_\mu\cdot\br + \phi_\mu(\br),\\
        &\phi_\mu(\br) =\frac{1}{\pi}\int d^2\br' \frac{\delta\theta(\br')}{\theta} \frac{\bq_\mu\cdot(\br-\br')}{|\br-\br'|^2}.\\
    \end{aligned}
\end{equation}
Note that $\bq_\mu$ is associated with the momentum scale of the moiré reciprocal lattice vectors. Putting all of this together, our expression for the moire potential in the presence of disorder takes the form
\begin{equation}
    \Delta(\br) = \frac{V}{2}e^{i\varphi}\sum_\mu e^{-i\bq_\mu\cdot \br- i\phi_\mu(\br)} + h.c. \label{eq:moire_potential}
\end{equation}

    \begin{figure*}[ht] 
    \centering
    \includegraphics[width=\textwidth]{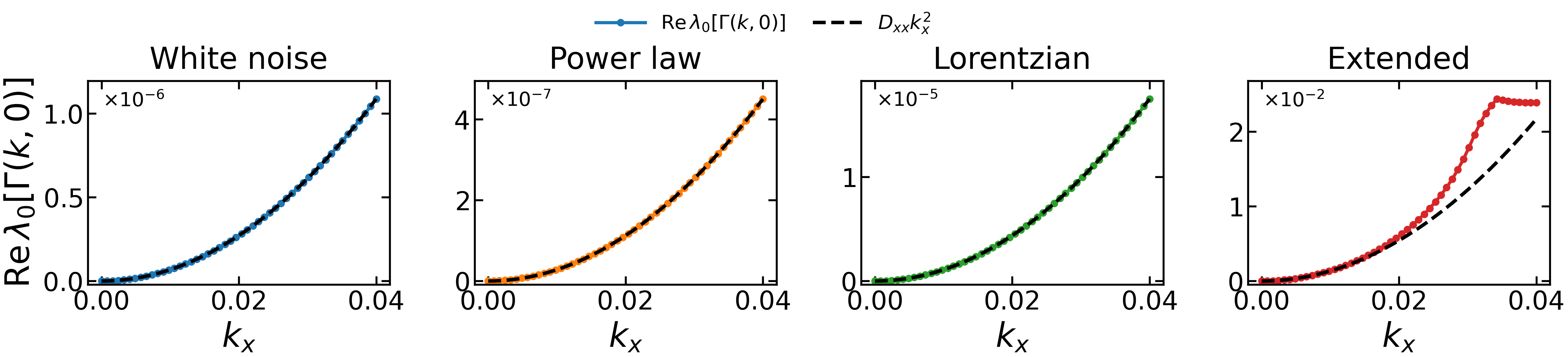} 
    \caption{Lowest (density) mode for the four different covariance kernels described earlier, calculated by discretising the Fermi surface into 720 angular patches. The strength of the trigonal warping is set to $\eta=0.03$. Data is connected by solid lines; quadratic fit is overlaid with dashes. Note the difference in scales across the models.}
    \label{fig:gamma}
\end{figure*}
\paragraph{Disorder averaging.} To characterise the effects of the twist-angle disorder, we now disorder average $\Delta(\br)$ under the correlation function $C_\theta(\br-\br')$. More explicitly, we seek to evaluate the functional integral $\int \mathcal{D}[\delta\theta]P[\delta\theta]e^{-S_0-S_\text{pot}}$, where
\begin{equation}
    P[\delta\theta]=\exp\left\{-\frac{1}{2}\delta\theta(\br)C_\theta^{-1}(\br-\br')\delta\theta(\br')\right\}
\end{equation}
We use the replica trick, $\braket{\ln Z}_{\delta\theta}=\lim_{N\to0}\frac{1}{N}(\braket{Z^N}_{\delta\theta}-1)$, to evaluate the replicated partition function $Z^N$ under disorder averaging. Both $S_0$, the free fermion part, and $S_\text{pot}$, the moiré potential part, should be therefore understood in the replica sense. We consider the situation when the twist-angle disorder is weak and the problem can be analysed perturbatively by the method of cumulant expansion. Up to second order in the cumulants, we approximately have $\braket{e^{-S_\text{pot}}}_{\delta\theta}\approx e^{-S_\text{dis}}$ where $S_\text{dis}=\braket{S_\text{pot}}_{\delta\theta}-\frac{1}{2}\left(\braket{S_\text{pot}^2}_{\delta\theta}-\braket{S_\text{pot}}_{\delta\theta}^2\right)\equiv S_\text{dis}^{(1)}+ S_\text{dis}^{(2)}$. It is straightforward to see that the first-order effect, $S_\text{dis}^{(1)}$, entails nothing more than putting fermions in a effective moiré potential defined by the mean twist angle. Thus the essential disorder physics is actually encoded in the second-order term,
\begin{equation}
    S_\text{dis}^{(2)}=-\sum_{a,b}\int d\tau d\tau' d^2\bm{r}d^2\bm{r}' g(\br,\br')\rho_a(\tau,\br)\rho_b(\tau',\br'),\label{eq:Sdis2}
\end{equation}
where $\rho_a(\tau,\br)={\psi}_a^{\dagger}(\tau,\br){\psi}_a(\tau,\br)$ is the replicated fermion density and, setting $\varphi=\pi/2$ in Eq.\eqref{eq:moire_potential},
  \begin{equation}
    \begin{aligned}
         g(\br,\br')=\frac{V^2}{4}\sum_{\mu\mu',s=\pm1}&s\cos\left(\bq_\mu\cdot \br-s\bq_{\mu'}\cdot\br'\right)\\
         &\times\left(V^{(2,s)}_{\mu\mu'}(\br-\br')-1\right).\label{eq:grR}
    \end{aligned}
  \end{equation}
The function $V$ is given by
\begin{equation}
     \begin{aligned}
         V^{(2,s)}_{\mu\mu'}(\br,\br')&=\braket{:e^{\pm i \phi_\mu(\br)}::e^{\mp is\phi_{\mu'}(\br')}:}\\
         &= e^{sG_{\mu\mu'}(\br-\br')}, ~~ s=\pm1.\\
     \end{aligned}\label{eq:V1V2}
 \end{equation} 
This term involves the disorder-induced correlation of $\phi_\mu(\br)$, defined as
$G_{\mu\mu'}(\br-\br')=\braket{\phi_\mu(\br)\phi_{\mu'}(\br')}_{\delta\theta}$. For generic kernel $S(\bk)$,
\begin{equation}
    \begin{aligned}
        G_{\mu\mu'}(\bR)&= \frac{1}{\pi^2\theta^2}\int d^2\bk e^{i\bk\cdot\bR}S(\bk)\frac{(\bq_\mu\cdot\bk)(\bq_{\mu'}\cdot\bk)}{|\bk|^4}
    \end{aligned}
\end{equation}
As in the conventional impurity scattering case, the disorder average introduces a correlation $g(\br,\br')$ between fermion density operators. 

Now we can, at last, make some qualitative statements about the effect of disorder. First of all, notice that the sign of $g(\br,\br')$ oscillates in space. This means that both positive and negative effective correlations will be induced between replicas. Similar effects have also been identified for twist-angle disorder using a phenomenological model \cite{paper1}. Because the correlation length of the disorder is much larger than the moire lattice scale, we can perform a Wigner transform and subsequently average over the centre-of-mass coordinate across a single moire unit cell (MUC) to arrive at the MUC-averaged coupling, which to first order in the Green's function is (see SM for details)
\begin{equation}
    \bar g(\delta\br) = \frac{V^2}{4}\sum_\mu G_{\mu\mu}(\delta\br)\cos(\bq_\mu\cdot \bR),
\end{equation}
where $\delta\br=\br-\br'$, and its Fourier transform is
\begin{equation}
    \begin{aligned}
        \bar g(\bk) = \frac{V^2}{8}\sum_\mu\left[ \tilde G_{\mu\mu}(\bk-\bq_\mu) + \tilde G_{\mu\mu}(\bk+\bq_\mu) \right];\\
        \tilde G_{\mu\mu'}(\bk) = \frac{1}{\theta^2}S(\bk)\frac{(\bq_\mu\cdot\bk)(\bq_{\mu'}\cdot\bk)}{k^4}
    \end{aligned} \label{eq:gk}
\end{equation}
When $S(\bk)=1$ (white noise), the functional form of $\bar g(\bk)$ is strongly reminiscent of what one finds in the presence of a static random magnetic field \cite{PhysRevB.49.16609}. One should not be so surprised by this; the relation between the displacement field and twist angle is exactly analogous to that between the vector potential and magnetic field. Indeed, processes for which $\bk\simeq \bq_\mu$ are strongly singular \footnote{To determine whether the singularity is a numerical problem, we introduce a small regularizer to the denominator of Eq. \ref{eq:gk} and then take it to zero. The physical quantities we later calculate, e.g., the diffusivity, converge as the regularizer is taken to zero. As a result, we conclude that the singularity is not obscuring the relevant physics.}. If we assume that scattering is taking place on the Fermi surface, this corresponds to the condition that $2k_F\simeq q_0$, where $q_0\equiv |\bq_\mu|$. The Fermi wave-vector $k_F$ is determined by filling factor. Therefore, one expects that the effects of disorder can depend strongly on filling.  

\paragraph{Non-linear sigma model.} The generating functional for a two-particle Green's function of the retarded-advanced type may be represented as a functional integral over a field $\psi\equiv \left( \psi_a^R~~~ \psi_a^A\right)^\dagger$, with associated metric $\Lambda = \text{diag}(1,-1)$. The partition function is $Z = \int \mathcal{D}[\psi] \exp\left[-S_0-S_\text{dis} \right]$, where
\begin{equation}
    S_0 = -i\sum_a\int d^2\br\left\{\bar\psi_a\left[(E_F-H_0)+\frac{\omega}{2}\Lambda + i0\Lambda\right]\psi_a \right\}
\end{equation}
is the free part, with $H_0$ representing the clean moire-band Hamiltonian, and 
\begin{equation}
    \begin{aligned}
        S_\text{dis} = \sum_{a,b}\sum_{\alpha,\beta}\int &d^2\br d^2\br' \left[ \bar\psi^\alpha_a(\br)\Lambda_\alpha\psi^\beta_b(\br')\right] \\
        &\times \bar g(\br-\br')\left[ \bar\psi^\beta_b(\br')\Lambda_\beta\psi^\alpha_a(\br)\right]
    \end{aligned}
\end{equation}
being the disorder-averaged quantity. Note that we have switched to a fixed-energy representation. Here, $a$ and $b$ denote replica indices and $\alpha,\beta$ correspond to retarded/advanced indices. The $+$ sign comes from rewriting the quartic term in the exchange channel. Because we are interested in the low-energy effective theory, we take the Fourier transform of $S_\text{dis}$ and project the disorder kernel onto the Fermi surface. To this end, consider only momenta $\bp$ and $\bp'$ on the Fermi surface. Let $\phi$ and $\phi'$ parameterise the positions of $\bp$ and $\bp'$, respectively, on the Fermi surface. We can now rewrite $S_\text{dis}$ as 
\begin{equation}
    \begin{aligned}
        S_\text{dis}& = \sum_{a,b}\sum_{\alpha,\beta}\int \frac{d^2\bk}{(2\pi)^2}\int d\mu(\phi)d\mu(\phi')\\
        &\times \left[ \bar\psi^\alpha_a\left(\bm{p}(\phi)-\frac{\bk}{2}\right)\Lambda_\alpha\psi^\beta_b\left(\bm{p}(\phi)+\frac{\bk}{2}\right)\right]\\
        &\times \bar g(\phi',\phi)\left[ \bar\psi^\beta_b\left(\bm{p}(\phi')+\frac{\bk}{2}\right)\Lambda_\beta\psi^\alpha_a\left(\bm{p}(\phi')-\frac{\bk}{2}\right)\right]
    \end{aligned}
\end{equation}
In the expression above, $d\mu(\phi)$ and $d\mu(\phi')$ are angular measures. Define now the bilinears
\begin{equation}
    \begin{aligned}
        X^{\alpha\beta,ab}_{m\ell} (\bk) &= \int d\mu(\phi) f^*_{m\ell}(\phi) \\
        &\times \bar\psi^\alpha_a\left(\bm{p}(\phi)-\frac{\bk}{2}\right)\Lambda_\alpha\psi^\beta_b\left(\bm{p}(\phi)+\frac{\bk}{2}\right)
    \end{aligned}
\end{equation}
where we label the eigenfunctions of the kernel $g(\phi,\phi')$ as $f_{m\ell}$, with corresponding eigenvalues $\lambda_{m\ell}$. The quantum number $m$ labels the $C_3$ representation and $\ell$ labels the state within that sector. Relative to the case of a static magnetic field studied in Ref. \cite{PhysRevB.49.16609}, the symmetry is reduced; here, different harmonics with the same $m$ quantum number can mix. For example, we use a simple trigonally-warped dispersion to capture the C$_3$ symmetry: 
\begin{equation}
    \varepsilon(\bk,\phi) = \frac{k^2}{2M} + \eta k^3 \cos 3\phi - E_F,
\end{equation}
where $M$ is the effective mass and $\eta$ is some warping parameter. We proceed to decouple the interaction with the aid of Hubbard-Stratonovich fields $Q_{m\ell}$, integrate out the fermions, and take the saddle-point approximation (see SM for details). We arrive at the following effective action for the Hubbard-Stratonovich fields: 
\begin{figure*}[t] 
    \centering
    \includegraphics[width=\textwidth]{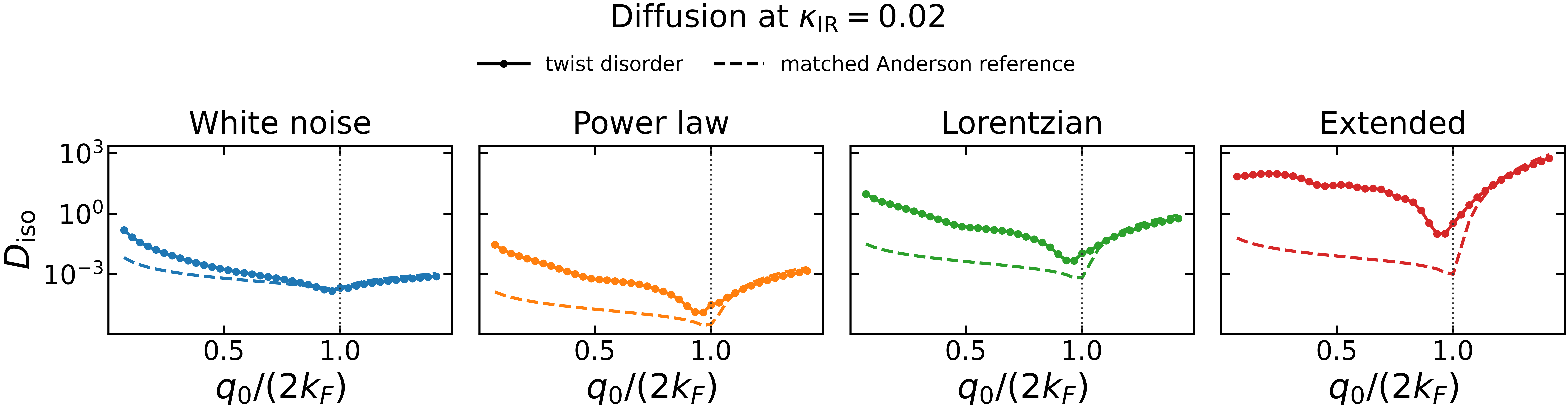} 
    \caption{Isotropic diffusivity, $D_\text{iso} = (D_{xx}+D_{yy})/2$, for the four covariance kernels described earlier, calculated by discretising the Fermi surface into 720 angular patches. The strength of the trigonal warping is set to $\eta=0.03$. The Anderson reference is a constant fermion scattering kernel matched to the mean rate of the twist-angle disorder scattering integrated across the Fermi surface.}
    \label{fig:diffusivity}
\end{figure*}
\begin{equation}
    \begin{aligned}
        S_\text{eff} &= S_0 +\int \frac{d^2\bk}{(2\pi)^2}\sum_{m\ell m'\ell'} \delta\tilde Q_{m\ell}(\bk)\Gamma_{m\ell,m'\ell'}\delta \tilde{Q}^\dagger_{m'\ell'}(\bk);
    \end{aligned}
\end{equation}
where 
\begin{equation}
    \Gamma_{m\ell,m'\ell'} = \lambda_{m\ell}\delta_{m\ell,m'\ell'}-\lambda_{m\ell}\lambda_{m'\ell'}\Pi_{m\ell,m'\ell'}(\bk;\omega) 
\end{equation}
and 
\begin{equation}
    \begin{aligned}
        \Pi_{m\ell,m'\ell'}(\bk;\omega) 
        \approx &\int d\mu(\phi) f^*_{m\ell}(\phi)f_{m'\ell'}(\phi) \\
        & \times G_q^A\left(\bp(\phi)-\frac{\bk}{2}, E_F-\frac{\omega}{2}\right) \\
        &\times G_q^R\left(\bp(\phi)+\frac{\bk}{2},E_F+\frac{\omega}{2} \right)
    \end{aligned}
\end{equation}
The soft modes are those for which, as $k\to 0$ and $\omega\to 0$, the corresponding eigenvalues of $\Gamma$ go to zero. We confirm numerically in Fig. \ref{fig:gamma} that the diffuson sector has the expected long-wavelength form $\Gamma(\bk,\omega)\sim -i\omega + D_{ij} k_i k_j + \dots$, with one conserved density mode and all other angular modes massive. Depending on the time-reversal symmetry of the microscopic Hamiltonian, the long-distance theory can be identified with one of several Altland-Zirnbauer universality classes. For the trigonal warping case presented in the numerics, the NLSM is in class A. 

\paragraph{Diffusion.} Diffusivities across four kernels are shown in Fig. \ref{fig:diffusivity}, alongside reference Anderson disorder (white noise coupled directly to the fermion density). The strength of the Anderson disorder is chosen such that the mean scattering rates of the two mechanisms are exactly matched. The horizontal axis tunes the ratio $q_0/2k_F$. We see that in all cases the diffusivity associated with twist-angle disorder is larger than that of the Anderson reference. This indicates that the twist-angle disorder is less effective at relaxing momentum, due to the added structure of the disorder-replicated kernel. In fact, the enhanced diffusivity is strongest in the `extended' angle disorder correlation kernel, which offers the closest match to experiment \cite{zeldov}. 

\paragraph{Discussion.} We considered here the case of two-dimensional fermions subject to twist-angle disorder. Our model is closely related to that of charged particles in a static magnetic field; in both cases, the replicated action can be mapped onto a NLSM. We show numerically that the resulting diffusivity is larger---by several orders of magnitude---than that which would result from mean-rate-matched white noise. This suggests that the twist disorder kernel is substantially less effective at relaxing momentum, owed in part to its intricate geometric structure. The fact that twist-angle disorder is less strongly localising than its Anderson counterpart has important consequences for understanding the stability of correlated phases, e.g., superconductivity and fractional quantum Hall, across moiré platforms.

{\it Acknowledgments.} We would like to thank Fengcheng Wu, Trithep Devakul, Sri Raghu, Pavel Nosov, Akshat Pandey, and Yi Huang for inspiring discussions. Y.M.W. is supported by a startup fund at Zhejiang University. N.S.T. is funded by Google Quantum AI.

   \bibliography{tbg}
   
\setcounter{equation}{0}
   \pagebreak
\onecolumngrid
\setcounter{figure}{0}
\setcounter{table}{0}
\setcounter{page}{1}%
\renewcommand{\theequation}{S\arabic{equation}}%
\renewcommand{\thefigure}{S\arabic{figure}}

\section{Supplement}
\subsection{Green's function derivation}
The key ingredient in evaluating disorder averages is the correlation (Green's) function for the disorder fields, defined as $G_{\mu\mu'}(\br-\br')=\braket{\phi_\mu(\br)\phi_{\mu'}(\br')}_{\delta\theta}$. It is given by
\begin{equation}
    G_{\mu\mu'}(\br-\br')= \frac{1}{\pi^2\theta^2}\int d^2\br'' d^2\br''' \frac{\bq_\mu\cdot(\br-\br'')}{|\br-\br''|^2} \frac{\bq_{\mu'}\cdot(\br'-\br''')}{|\br'-\br'''|^2} \langle \delta\theta(\br'')\delta\theta(\br''')\rangle  
\end{equation}
For generic kernel $S(\bk)$, we perform the integrals to arrive at
\begin{equation}
    \begin{aligned}
        G_{\mu\mu'}(\bR)&= \frac{1}{\pi^2\theta^2}\int d^2\bk e^{i\bk\cdot\bR}S(\bk)\frac{(\bq_\mu\cdot\bk)(\bq_{\mu'}\cdot\bk)}{|\bk|^4}
    \end{aligned}
\end{equation}

\begin{figure*}[t] 
    \centering
    \includegraphics[width=\textwidth]{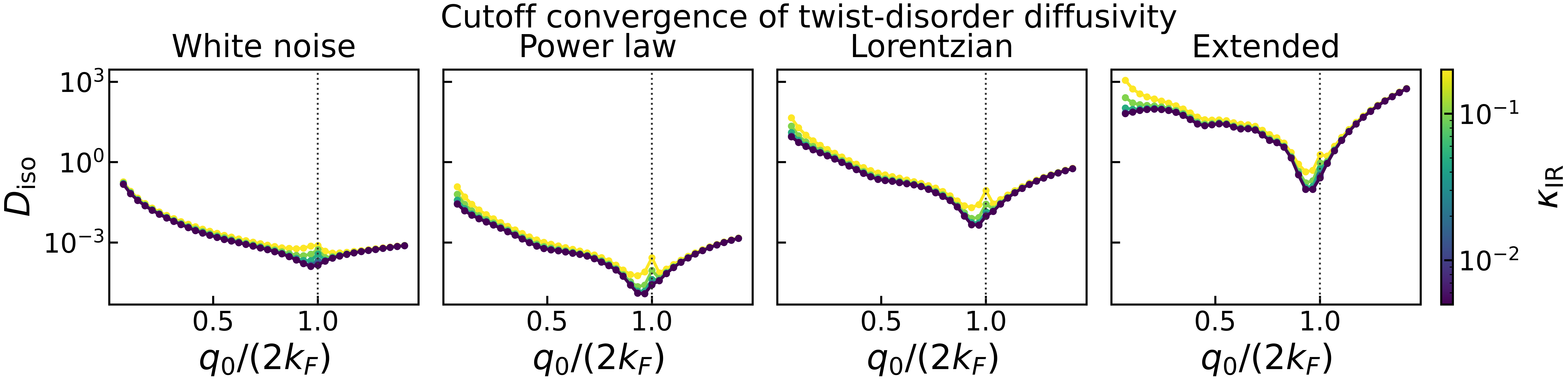} 
    \caption{Isotropic diffusivity calculated for the four covariance kernels described earlier, sweeping the small regularizer $\kappa_\text{IR}$. We find that the calculation converges as we take $\kappa_\text{IR}\to 0$. The strength of the trigonal warping is set to $\eta=0.03$}
    \label{fig:regularization}
\end{figure*}

\paragraph{Derivation.}
Start with
\begin{align}
  G_{\mu\mu'}(\br-\br')&= \frac{1}{\pi^2\theta^2}\int d^2\br'' d^2\br''' \frac{\bq_\mu\cdot(\br-\br'')}{|\br-\br''|^2} \frac{\bq_{\mu'}\cdot(\br'-\br''')}{|\br'-\br'''|^2} C_\theta(\br''-\br''') 
\end{align}
Define $\bx = \br-\br''$ and $\bm{y}=\br'-\br'''$. Furthermore, define $K_{\bq}(\bx)=(\bq\cdot\bx)/x^2$. We can now rewrite the integral using convolutions:
\begin{equation}
    \begin{aligned}
        G_{\mu\mu'} &=\frac{1}{\pi^2\theta^2}\int d^2\bx d^2\bm{y} ~\frac{\bq_\mu\cdot \bx}{x^2}\frac{\bq_{\mu'}\cdot \bm{y}}{y^2} C_\theta(\bR-\bx+\bm{y})\\
        &=\frac{1}{\pi^2\theta^2}\int d^2\bx d^2\bm{y} ~K_{\bq_\mu}(\bx) K_{\bq_{\mu'}}(\bm{y})C_\theta(\bR-\bx+\bm{y})\\
        &= \frac{1}{\pi^2\theta^2}(K_{\bq_\mu}*K_{\bq_{\mu'}}*C)(\bR)\\
        &= \frac{1}{\pi^2\theta^2}\int \frac{d^2\bk}{(2\pi)^2}e^{i\bk\cdot\bR}\tilde{K}_{\bq_\mu}(\bk) \tilde{K}_{\bq_{\mu'}}(-\bk)S(\bk)
    \end{aligned}
\end{equation}
The Fourier transform of $K_{\bq}$ is
\begin{equation}
    \begin{aligned}
        \tilde{K}_{\bq}(\bk) &= \int d^2\bx e^{-i\bk\cdot\bx}\frac{\bq\cdot\bx}{x^2}\\
        &= i\bq\cdot\nabla_{\bk}\int d^2\bx \frac{e^{-i\bk\cdot\bx}}{x^2}\\
        &=-i2\pi\bq\cdot\nabla_{\bk} \log|\bk|\\
        &=-i2\pi \frac{\bq\cdot\bk}{k^2}
    \end{aligned}
\end{equation}
Thus,
\begin{equation}
    \begin{aligned}
        G_{\mu\mu'}(\bR) &= \frac{1}{\pi^2\theta^2}\int d^2\bk e^{i\bk\cdot\bR}S(\bk)\frac{(\bq_\mu\cdot\bk)(\bq_{\mu'}\cdot\bk)}{|\bk|^4}\\
        &= -\frac{1}{\pi^2\theta^2}(\bq_\mu\cdot\nabla_{\bR})(\bq_{\mu'}\cdot\nabla_{\bR})\int d^2\bk e^{i\bk\cdot\bR}\frac{S(\bk)}{|k|^4}
    \end{aligned}\label{eq:G_double_nabla}
\end{equation}
To calculate $S_\text{dis}^{(2)}$, we also need the following correlation function:
\begin{equation}
     \begin{aligned}
         V^{(2,s)}_{\mu\mu'}(\br,\br')&=\braket{:e^{\pm i \phi_\mu(\br)}::e^{\mp is\phi_{\mu'}(\br')}:}\\
         &= e^{sG_{\mu\mu'}(\br-\br')}, ~~ s=\pm1.\\
     \end{aligned}\label{eq:V1V2}
 \end{equation} 
In terms of $V_{\mu\mu'}^{(2,s)}$, it is straightforward to see
\begin{equation}
    S_\text{dis}^{(2)}=-\int d\tau d\tau'\sum_{a,b}  d^2\bm{r}d^2\bm{r}' g(\br,\br')\rho_a(\tau,\br)\rho_b(\tau',\br')\label{eq:Sdis2}
\end{equation}
where $\rho_a(\tau,\br)={\psi}_a^{\dagger}(\tau,\br){\psi}_a(\tau,\br)$ is the replicated fermion density and 
  \begin{equation}
    \begin{aligned}
         g(\br,\br')&=\frac{V^2}{4}\sum_{\mu\mu',s=\pm1}s\cos\left(\bq_\mu\cdot \br-s\bq_{\mu'}\cdot\br'\right)\times\left(V^{(2,s)}_{\mu\mu'}(\br-\br')-1\right).\label{eq:grR}
    \end{aligned}
  \end{equation}
In arriving at this expression, we have taken $\varphi=\pi/2$ in the clean expression for the moiré potential.
We see that Eqs.\eqref{eq:Sdis2} and \eqref{eq:grR} are the second order effects from the twist angle disorder. As in the conventional impurity scattering case, the disorder average introduces a correlation $g(\br,\br')$ between fermion density operators. Now we can, at last, make some qualitative statements about the effect of disorder. First of all, notice that the sign of $g(\br,\br')$ oscillates in space. This means that both positive and negative effective correlations will be induced between replicas. To make analytic progress, we expand to first order in $G$ to get
\begin{equation}
\begin{aligned}
g(\br,\br')&=\frac{V^2}{4}\sum_{\mu\mu'}\sum_{s=\pm1}s\cos\left(\bq_\mu\cdot \br-s\bq_{\mu'}\cdot\br'\right)\times sG_{\mu\mu'}(\bR)\\
&= \frac{V^2}{4}\sum_{\mu\mu'}G_{\mu\mu'}(\bR)\sum_{s=\pm 1} \cos\left(\bq_\mu\cdot \br-s\bq_{\mu'}\cdot\br'\right)\\
&= \frac{V^2}{2}\sum_{\mu\mu'}G_{\mu\mu'}(\bR) \cos(\bq_\mu\cdot \br)\cos(\bq_{\mu'}\cdot \br')
\end{aligned}
\end{equation}
We will now express this quantity in terms of the centre of mass coordinate, $\bR_\text{CM} = (\br+\br')/2$, and the relative coordinate, $\bR = \br-\br'$:
\begin{equation}
    \begin{aligned}
        g(\bR_\text{cm}, \bR) &\approx \frac{V^2}{2}\sum_{\mu\mu'}G_{\mu\mu'}(\bR)\times\cos\left(\bq_{\mu}\cdot\left[ \bR_\text{cm}+\frac{\bR}{2}\right]\right)\cos\left(\bq_{\mu'}\cdot\left[ \bR_\text{cm}-\frac{\bR}{2}\right]\right)\\
        &= \frac{V^2}{4}\sum_{\mu\mu'}G_{\mu\mu'}(\bR)\times \left\{\cos\left([\bq_\mu+\bq_{\mu'}]\cdot\bR_\text{cm}+ [\bq_\mu-\bq_{\mu'}]\cdot\frac{\bR}{2}\right)+\cos\left([\bq_\mu-\bq_{\mu'}]\cdot\bR_\text{cm}+ [\bq_\mu+\bq_{\mu'}]\cdot\frac{\bR}{2}\right)\right\}
    \end{aligned}
\end{equation}
Because the correlation length of the disorder is much larger than the moire lattice scale, we can average over the centre-of-mass coordinate $\bR_\text{cm}$ over a single moire unit cell. The key orthogonality relation here is 
\begin{equation}
    \frac{1}{A_\text{muc}}\int_\text{muc}d^2\bR_\text{cm} \cos(\bK\cdot\bR_\text{cm}+\delta) = \begin{cases}
        \cos\delta, &\bK=0 \\
        0, &\text{otherwise}
    \end{cases}
\end{equation}
where $\int_\text{muc}$ is the integral over the moire unit cell and $A_\text{muc}$ is its area. It follows that the averaged coupling is 
\begin{equation}
    \bar g(\bR) = \frac{V^2}{4}\sum_\mu G_{\mu\mu}(\bR)\cos(\bq_\mu\cdot \bR)
\end{equation}
Writing $\cos(\bq_\mu\cdot \bR) = \frac{1}{2}\left(e^{i\bq_\mu\cdot\bR}+e^{-i\bq_\mu\cdot\bR}\right)$, we see that the Fourier transform is given by
\begin{equation}
    \begin{aligned}
        \bar g(\bk) &= \frac{V^2}{8}\sum_\mu\left[ \tilde G_{\mu\mu}(\bk-\bq_\mu) + \tilde G_{\mu\mu}(\bk+\bq_\mu) \right]
    \end{aligned}
\end{equation}
where
\begin{equation}
    \tilde G_{\mu\mu'}(\bk) = \frac{1}{\theta^2}S(\bk)\frac{(\bq_\mu\cdot\bk)(\bq_{\mu'}\cdot\bk)}{k^4}
\end{equation}
Plugging this in above, we have
\begin{equation}
    \bar g(\bm{k}) = \frac{V^2}{8\theta^2} \sum_\mu \left(S(\bm{k}-\bm{q}_\mu)\frac{[\bm{q}_\mu\cdot(\bm{k}-\bm{q}_\mu)]^2}{|\bm{k}-\bm{q}_\mu|^4}+S(\bm{k}+\bm{q}_\mu)\frac{[\bm{q}_\mu\cdot(\bm{k}+\bm{q}_\mu)]^2}{|\bm{k}+\bm{q}_\mu|^4}\right) \label{eq:gk}
\end{equation}
We add to the denominator a small regularizer, $\kappa_\text{IR}$, to check convergence. Specifically, we take 
\begin{equation}
    |\bk\pm \bq_\mu|^4 \to \left(|\bk\pm \bq_\mu|^2 + \kappa^2_\text{IR} \right)^2
\end{equation}
We find that the numerics remain well-controlled even as $\kappa_\text{IR}\to 0$, see Fig. \ref{fig:regularization}.

\subsection{Non-linear sigma model}
The generating functional for a two-particle Green's function of the retarded-advanced type may be represented as a functional integral over a field 
\begin{equation}
    \psi\equiv\begin{pmatrix}
        \psi_a^R\\
        \psi_a^A
    \end{pmatrix}
\end{equation}
with associated metric 
\begin{equation}
    \Lambda = \text{diag}(1,-1).
\end{equation}
The partition function is
\begin{equation}
    Z = \int \mathcal{D}[\psi] \exp\left[-S_0-S_\text{dis} \right],
\end{equation}
where
\begin{equation}
    S_0 = -i\sum_a \int d^2\br\left\{\bar\psi_a\left[(E-H_0)+\frac{\omega}{2}\Lambda + i0\Lambda\right]\psi_a \right\}
\end{equation}
is the free part, with $H_0$ representing the clean moire-band Hamiltonian, and 
\begin{equation}
    S_\text{dis} = \sum_{a,b}\sum_{\alpha,\beta} \int d^2\br d^2\br' \left[ \bar\psi^\alpha_a(\br)\Lambda_\alpha\psi^\beta_b(\br')\right] \bar g(\br-\br')\left[ \bar\psi^\beta_b(\br')\Lambda_\beta\psi^\alpha_a(\br)\right]
\end{equation}
is the disorder-averaged quantity. Here, $a$ and $b$ denote replica indices and $\alpha,\beta$ correspond to retarded/advanced indices. The $+$ sign comes from rewriting the quartic term in the exchange channel:
\begin{equation}
    (\bar\psi^\alpha_a \psi^\alpha_a)(\bar\psi^\beta_b \psi^\beta_b) = -(\bar\psi^\alpha_a \psi^\beta_b)(\bar\psi^\beta_b \psi^\alpha_a)
\end{equation}
Next we take the Fourier transform of $S_\text{dis}$ and suggestively write the interaction in the exchange channel:
\begin{equation}
    S_\text{dis} =  \sum_{a,b}\sum_{\alpha,\beta}\int\frac{ d^2\bm{p}}{(2\pi)^2} \frac{ d^2\bm{p}'}{(2\pi)^2}\frac{ d^2\bm{k}}{(2\pi)^2}  \left[ \bar\psi^\alpha_a\left(\bm{p}-\bm{k}/2\right)\Lambda_\alpha\psi^\beta_b\left(\bm{p}+\bm{k}/2\right)\right] \bar g(\bm{p}'-\bm{p})\left[ \bar\psi^\beta_b\left(\bm{p}'+\bm{k}/2\right)\Lambda_\beta\psi^\alpha_a\left(\bm{p}'-\bm{k}/2\right)\right]
\end{equation}
So far we have considered the problem in full generality. Because we are interested in the low energy effective theory, we will now project the disorder kernel onto the Fermi surface. To this end, consider only momenta $\bp$ and $\bp'$ on the Fermi surface. Let $\phi$ and $\phi'$ parameterize the positions of $\bp$ and $\bp'$, respectively, on the Fermi surface. We can now rewrite $S_\text{dis}$ as 
\begin{equation}
    \begin{aligned}
        S_\text{dis} = &\sum_{a,b}\sum_{\alpha,\beta} \int \frac{d^2\bk}{(2\pi)^2}\int d\mu(\phi)d\mu(\phi')\\
        &\times \left[ \bar\psi^\alpha_a\left(\bm{p}(\phi)-\frac{\bk}{2}\right)\Lambda_\alpha\psi^\beta_b\left(\bm{p}(\phi)+\frac{\bk}{2}\right)\right] \bar g(\phi',\phi)\left[ \bar\psi^\beta_b\left(\bm{p}(\phi')+\frac{\bk}{2}\right)\Lambda_\beta\psi^\alpha_a\left(\bm{p}(\phi')-\frac{\bk}{2}\right)\right]
    \end{aligned}
\end{equation}
In the expression above, $d\mu(\phi)$ and $d\mu(\phi')$ are angular measures. Define now the bilinears
\begin{equation}
    X^{\alpha\beta,ab}_{m\ell} (\bk) = \int d\mu(\phi) f^*_{m\ell}(\phi) \bar\psi^\alpha_a\left(\bm{p}(\phi)-\frac{\bk}{2}\right)\Lambda_\alpha\psi^\beta_b\left(\bm{p}(\phi)+\frac{\bk}{2}\right)
\end{equation}
where we can label the eigenfunctions of the kernel $g(\phi,\phi')$ using the labels $m$ (angular momentum modulo 3) and $\ell$ (labels the different eigenvectors inside the $m$ sector)
\begin{equation}
    f_{m\ell}(\phi) = \sum_{s\in\mathbb{Z}} u_s^{(m\ell)} e^{i(m+3s)\phi}; \qquad m=0,1,2
\end{equation}
Relative to the case of a static magnetic field studied in Ref. \cite{PhysRevB.49.16609}, the symmetry is reduced; here, different harmonics with the same $m$ quantum number can mix. We may now rewrite the disordered term as a sum over quasi-angular momentum sectors:
\begin{equation}
    S_\text{dis} = \sum_{m,\ell}\lambda_{m\ell} \int \frac{d^2\bk}{(2\pi)^2}X^{\alpha\beta,ab}_{m\ell}(\bk) X^{\beta\alpha,ba}_{m\ell}(-\bk)
\end{equation}
The interaction term may be decoupled with the aid of Hubbard-Stratonovich fields $Q_{m\ell}\equiv Q^{ab,\alpha\beta}_{m\ell}$ and $ \left[Q_{m\ell}^{ab,\alpha\beta}(\bk) \right]^\dagger = Q^{ba,\beta\alpha}_{m\ell}(-\bk)$. To decouple the interaction, we first shift the bilinears by
\begin{equation}
    S_\text{coupled} = -i\sum_{m\ell}\lambda_{m\ell}\int \frac{d^2\bk}{(2\pi)^2}\left[X_{m\ell}(\bk) Q^\dagger_{m\ell}(\bk) + Q_{m\ell}(\bk)X^\dagger_{m\ell}(\bk)\right]
\end{equation}
and then multiply by the ``fat unity''. This brings the partition function into the form 
\begin{equation}
    Z = \int \mathcal{D}[Q]\mathcal{D}[\bar\psi,\psi]\exp \left\{-S[Q,\bar\psi,\psi] \right\},
\end{equation}
where
\begin{equation}
    \begin{aligned}
        S[Q,\bar\psi,&\psi] = \int \frac{d^2\bk}{(2\pi)^2}\sum_{m,\ell}\lambda_{m\ell}~ Q_{m\ell}(\bk)Q^\dagger_{m\ell}(\bk)\\
        &+\sum_{ab,\alpha\beta} \int \frac{d^2\bk}{(2\pi)^2}\frac{d^2\bk'}{(2\pi)^2} \int d\mu(\phi)~ \bar\psi_a^\alpha(\bp(\phi) + \bk)\Bigg[\delta_{ab}\delta_{\alpha\beta}(2\pi)^2\delta^2(\bk-\bk')G_{0,\alpha}^{-1}(\bp(\phi)+\bk) \\
        &\qquad \qquad\qquad \qquad \qquad \qquad \qquad \qquad  + i\Lambda_\alpha\sum_{m,\ell} \lambda_{m\ell}f^*_{m\ell}(\phi)  Q^\dagger_{m\ell}(\bk'-\bk)\Bigg]\psi^\beta_b(\bp(\phi)+\bk')
    \end{aligned}
\end{equation}
Upon integrating out fermions, the resulting partition function takes the form
\begin{equation}
    Z = \int \mathcal{D}[ Q] \exp\left\{-S_\text{eff}[Q]\right\}
\end{equation}
with effective action of the $Q$ fields given by
\begin{equation}
    S_\text{eff}[Q] = -\text{Tr}\ln G^{-1}+  \int \frac{d^2\bk}{(2\pi)^2} \sum_{m,\ell}\lambda_{m\ell}~ Q_{m\ell}(\bk) Q^\dagger_{m\ell}(\bk)
\end{equation}
In the expression above, the Green's function is
\begin{equation}
    \left[G^{-1}(\phi; \bk,\bk')\right]^{\alpha\beta}_{ab} = \delta_{ab}\delta_{\alpha\beta}(2\pi)^2 \delta^2(\bk-\bk')G_{0,\alpha}^{-1}( \bp(\phi)+\bk)+i\Lambda_\alpha\sum_{m,\ell} \lambda_{m\ell}f^*_{m\ell}(\phi)  Q^\dagger_{m\ell}(\bk'-\bk) \label{eq:G_NLSM}
\end{equation}
with
\begin{equation}
    G^{-1}_{0,\alpha} = E-H_0 + \Lambda_\alpha\left(\frac{\omega}{2}+i0 \right)
\end{equation}
The saddle point of the action, $Q_\text{saddle}\equiv q$, satisfies
\begin{equation}
    \frac{\delta S_\text{eff}}{\delta q^{ab,\alpha\beta}_{m\ell}(\bk)} = -i\Lambda_\alpha \lambda_{m\ell}  \int d\mu(\phi) f^*_{m\ell}(\phi)\int\frac{d^2\bk'}{(2\pi)^2}~ G^{ba}_{\beta\alpha}(\phi;\bk',\bk'-\bk) + \lambda_{m\ell}q_{m\ell}^{ba,\beta\alpha}(-\bk)=0
\end{equation}
Solving this equation for $q$ gives the self-consistent Born approximation,
\begin{equation}
q^{ab,\alpha\beta}_{m\ell}(\bk) = i\Lambda_\alpha\int d\mu(\phi)f^*_{m\ell}(\phi) \int\frac{d^2\bk'}{(2\pi)^2} G_{ab}^{\alpha\beta}(\phi;\bk',\bk'-\bk)
\end{equation}
At the level of the saddle-point approximation, one typically assumes that the field $q$ is perfectly homogeneous and furthermore that it is diagonal in both replica and retarded/advanced index: 
\begin{equation}
    q_{m\ell}^{ab,\alpha\beta}(\bk) \equiv \delta_{ab}\delta_{\alpha\beta}(2\pi)^2\delta^2(\bk)\delta_{m,0}q^\alpha_{\ell}
\end{equation}
The Green's function evaluated at this saddle point, $G_q$, is
\begin{equation}
    \left[G_q^{-1}(\phi; \bk,\bk')\right]^{\alpha\beta}_{ab} = \delta_{ab}\delta_{\alpha\beta}(2\pi)^2 \delta^2(\bk-\bk')\left[G_{0,\alpha}^{-1}( \bp(\phi)+\bk)+i\Lambda_\alpha\sum_{\ell} \lambda_{0\ell}f^*_{0\ell}(\phi)  q^\alpha_{0\ell}\right]
\end{equation}
Substituting $G_q$ into the saddle point equation, we arrive at
\begin{equation}
    q^\alpha_\ell = i \Lambda_\alpha\int d\mu(\phi) \int\frac{d^2\bk'}{(2\pi)^2}\frac{f^*_{0\ell}(\phi)}{G^{-1}_{0,\alpha}(\bp(\phi)+\bk') + i\Lambda_\alpha\sum_{\ell'}\lambda_{0\ell'}f^*_{0,\ell'}(\phi)q^\alpha_{0\ell'}} 
\end{equation}
This means that the saddle-point energy is given by
\begin{equation}
    \Sigma_\text{sp} = -\frac{i}{2\tau(\phi)}\Lambda; \qquad \frac{1}{2\tau(\phi)} = \sum_\ell \lambda_{0\ell}f^*_{0\ell}(\phi)q_{0\ell}
\end{equation}
We now proceed to expand the action around the saddle point:
\begin{equation}
    \begin{aligned}
        S_\text{eff} = S_0 +  \int \frac{d^2\bk}{(2\pi)^2}\Bigg\{\int d\mu(\phi)&\sum_{m\ell}\sum_{m'\ell'}\left[  \lambda_{m\ell}\lambda_{m'\ell'}f^*_{m\ell}(\phi)f_{m'\ell'}(\phi)\right]\times \text{Tr}\left[ G_q(\bp(\phi))\Lambda \delta Q_{m\ell}(\bk) G_q(\bp(\phi)+\bk)\Lambda \delta  Q^\dagger_{m'\ell'}(\bk)\right] \\
        &-\sum_{m\ell}\lambda_{m\ell}~\text{Tr}\left[ \delta Q_{m\ell}(\bk)\delta Q^\dagger_{m\ell}(\bk)\right]\Bigg\}
    \end{aligned}
\end{equation}
As earlier, we have $\delta Q^\dagger(\bk) = \delta Q(-\bk)$. The integral over the Fermi surface is finite only for the products $G^R G^A$, which are generated by the off-diagonal components of $\delta Q$, which we from here on out denote $\delta\tilde Q$. In the limit of small $k$ and $\omega$, we define  
\begin{equation}
    \begin{aligned}
        \Pi_{m\ell,m'\ell'}(\bk;\omega) 
        &\approx \int d\mu(\phi) f^*_{m\ell}(\phi)f_{m'\ell'}(\phi) G_q^A\left(\bp(\phi)-\frac{\bk}{2}, E-\frac{\omega}{2}\right) G_q^R\left(\bp(\phi)+\frac{\bk}{2},E+\frac{\omega}{2} \right)
    \end{aligned}
\end{equation}
The effective action can therefore be written as
\begin{equation}
    S_\text{eff} = S_0 + \int \frac{d^2\bk}{(2\pi)^2}\sum_{m\ell m'\ell'} \delta\tilde Q_{m\ell}(\bk)\Gamma_{m\ell,m'\ell'}\delta \tilde{Q}^\dagger_{m'\ell'}(\bk)
\end{equation}
where
\begin{equation}
    \Gamma_{m\ell,m'\ell'} = \lambda_{m\ell}\delta_{m\ell,m'\ell'}-\lambda_{m\ell}\lambda_{m'\ell'}\Pi_{m\ell,m'\ell'}(\bk;\omega) 
\end{equation}
The soft modes are those for which, as $k\to 0$ and $\omega\to 0$, the corresponding eigenvalues of $\Gamma$ go to zero. We now proceed with approximating $\Pi$. Moving forward, we consolidate the labels $m$ and $\ell$ into a single channel label, $A$. In this notation,
\begin{equation}
    \Gamma_{A,B} = \lambda_A \delta_{A,B} - \lambda_A \lambda_B \Pi_{A,B}
\end{equation}
Earlier, we were a bit imprecise with what we meant by $\bp(\phi)$. Define it carefully as 
\begin{equation}
    \bp(\phi) = \bp_F(\phi)+\delta p_\perp \bm{\hat{n}}(\phi)
\end{equation}
Furthermore, define $\xi \equiv \varepsilon_{\bp}-E_F$. To linear order near the Fermi surface, $\xi \simeq v_F(\phi) \delta p_\perp$. The momentum measure, which we had initially approximated just using $d\mu(\phi)$, is actually 
\begin{equation}
    \int \frac{d^2\bp}{(2\pi)^2} =\nu \int d\mu(\phi) \int d\xi 
\end{equation}
If we assume a trigonally-warped Fermi surface, then 
\begin{equation}
    d\mu(\phi) \simeq \frac{d\phi}{2\pi}\left(1 + \delta \cos 3\phi\right)
\end{equation}
We assume that the velocity anisotropy does not significantly modify the geometric arc-length factor. Right at the homogeneous saddle, the retarded/advanced Green's functions are
\begin{equation}
    \begin{aligned}
        G^{R/A}_q(\phi,\xi) = \frac{1}{E-\varepsilon_{\bm{p}} \pm \frac{i}{2\tau_\phi}} = \frac{1}{-\xi \pm \frac{i}{2\tau_\phi}}
    \end{aligned}
\end{equation}
What we are actually interested in computing is
\begin{equation}
    G_q^A\left(\bp(\phi)-\frac{\bk}{2}, E-\frac{\omega}{2}\right) G_q^R\left(\bp(\phi)+\frac{\bk}{2},E+\frac{\omega}{2}\right)
\end{equation}
To this end, we linearize the dispersion as
\begin{equation}
    \varepsilon_{\bp\pm \bk/2}\simeq \varepsilon_{\bp} \pm \frac{1}{2}\bk \cdot \bm{v}_F(\phi)
\end{equation}
Thus the product is
\begin{equation}
    G^RG^A = \frac{1}{\left[-\xi + \frac{\omega - \bm{v}_F(\phi)\cdot\bk}{2} + \frac{i}{2\tau_\phi} \right]\left[-\xi - \frac{\omega - \bm{v}_F(\phi)\cdot\bk}{2} - \frac{i}{2\tau_\phi} \right]}
\end{equation}
Performing the integral over $\xi$ gives
\begin{equation}
    \int d\xi G^R G^A = \frac{2\pi \tau_\phi}{1-i\left[\omega-\bm{v}_F(\phi)\cdot\bk \right]\tau_\phi}
\end{equation}
Finally, we expand in small $k$ and $\omega$:
\begin{equation}
    \int d\xi G^R G^A \simeq 2\pi \tau_\phi \times \left[1+i\omega\tau_\phi - i\tau_\phi \bm{v}_F(\phi)\cdot\bk - \tau_\phi^2(\bm{v}_F(\phi)\cdot\bk)^2 + \cdots \right]
\end{equation}
Schematically,
\begin{equation}
    \Pi_{AB} = \Pi_{AB}^{(0)}+i\omega \Pi_{AB}^{(\omega)} - ik_i \Pi_{AB}^{(i)}-k_ik_j \Pi_{AB}^{(ij)},
\end{equation}
where
\begin{equation}
    \begin{aligned}
        \Pi_{AB}^{(0)} &= 2\pi \nu \int d\mu(\phi) f^*_A(\phi) f_B(\phi) \tau_\phi\\
        \Pi_{AB}^{(\omega)} &= 2\pi \nu \int d\mu(\phi) f^*_A(\phi) f_B(\phi) \tau^2_\phi\\
        \Pi_{AB}^{(i)} &= 2\pi \nu \int d\mu(\phi) f^*_A(\phi) f_B(\phi) \tau^2_\phi v_{i} (\phi)\\
        \Pi_{AB}^{(ij)} &= 2\pi \nu \int d\mu(\phi) f^*_A(\phi) f_B(\phi) \tau^3_\phi v_{i}(\phi) v_j (\phi)
    \end{aligned}
\end{equation}

\subsection{Numerics}
Begin with the trigonally-warped dispersion
\begin{equation}
    \varepsilon(\bk,\phi) = \frac{k^2}{2m}+\eta k^3 \cos 3\phi -\mu 
\end{equation}
We wish to discretise the Fermi surface into $N_\phi$ patches. For each of $N_\phi$ angles, solve
\begin{equation}
    \varepsilon(k_F(\phi_i),\phi_i) = 0
\end{equation}
The momenta along the Fermi surface are therefore given by:
\begin{equation}
    \bp_i = \bp_F(\phi_i) = k_F(\phi_i)\times (\cos\phi_i,\sin\phi_i)
\end{equation}
We can calculate the Fermi velocity at each point by taking the gradient with respect to $\bk$. Explicitly, the code uses
\begin{equation}
    \begin{aligned}
    \bm v_F = v_r\hat{\bm{e}}_r + v_\phi \hat{\bm{e}}_\phi
    \end{aligned}
\end{equation}
for 
\begin{equation}
        v_r = \frac{k_F}{m} + 3\eta k_F^2 \cos 3\phi; \qquad v_\phi = -3\eta k_F^2\sin 3\phi
\end{equation}
All of this allows us to properly define an integral over the Fermi surface. If we wish to integrate some function $F$ over the Fermi surface, we write
\begin{equation}
    \int \frac{d^2\bp}{(2\pi)^2}\delta(\varepsilon_{\bp}-E_F) F(\bp) \propto \oint_{\text{FS}} \frac{ds}{|\bm{v}_F|} F(s)
\end{equation}
For a surface parameterised by $\phi$,
\begin{equation}
    ds = \sqrt{k^2_F + \left( \frac{dk_F}{d\phi}\right)^2} d\phi 
\end{equation}
The code therefore constructs the weights 
\begin{equation}
    \tilde w_i \equiv \frac{1}{|\bm{v}_F(\phi_i)|}\sqrt{k^2_F(\phi_i) + \left( \frac{dk_F}{d\phi_i}\right)^2},
\end{equation}
then normalizes them such that 
\begin{equation}
    w_i = \frac{\tilde w_i}{\sum_j \tilde w_j}; \qquad \sum_i w_i = 1
\end{equation}
Consequently, the integral over the Fermi surface becomes
\begin{equation}
    \int d\mu(\phi) F(\phi) \to \sum_i w_i F_i
\end{equation}
For each momentum transfer $\bk = \bp_j - \bp_i$ on the Fermi surface, the code evaluates the IR-regularized $\bar g(\bk)$. The code defines the explicitly-symmetrised quantity
\begin{equation}
    W_{ij} = 2\pi \nu \times \frac{1}{2}\left[\bar g(\bp_j-\bp_i) + \bar g(\bp_i-\bp_j) \right]
\end{equation}
Physically, $W_{ij}$ is the rate kernel for scattering a particle from angle $\phi_i$ to $\phi_j$. The total elastic scattering rate out of patch $i$ is
\begin{equation}
    R_i = \sum_j w_j W_{ij},
\end{equation}
with associated scattering time $\tau_i =R_i^{-1}$. The next step is to construction the collision operator, $L$, defined as 
\begin{equation}
    L_{ij} = R_i \delta_{ij} - w_j W_{ij}
\end{equation}
We also construct its symmetrized counterpart,
\begin{equation}
    L^\text{sym}_{ij} = R_i \delta_{ij} - \sqrt{w_iw_j} W_{ij},
\end{equation}
which enters into the expression for $\Gamma$:
\begin{equation}
    \Gamma(\bk,\omega) = L^\text{sym}-i\omega I + i~\text{diag}(\bk\cdot\bm{v}_F),
\end{equation}
which we diagonalize at $\omega=0$ and for a range of small $k$ in order to determine the spectrum. At small $k$, the term proportional to $\bk\cdot\bm{v}_F$ couples the density mode to the massive angular modes; this is how diffusion arises. The diffusion constant can be extracting by fitting the small-$k$ behaviour to a quadratic dispersion. 

\end{document}